\documentclass[12pt,reqno]{article}
\usepackage{amsmath}

\allowdisplaybreaks
\numberwithin{equation}{section}

\newcommand{\al}{\alpha}
\newcommand{\ad}{\dot{\alpha}}
\newcommand{\be}{\beta}

\newcommand{\ga}{\gamma}

\newcommand{\de}{\delta}

\newcommand{\si}{\sigma}
\newcommand{\bsi}{\bar{\sigma}}
\newcommand{\bpsi}{\bar{\psi}}
\newcommand{\half}{\tfrac{1}{2}}
\newcommand{\ihalf}{\tfrac{\mathrm{i}}{2}}
\newcommand{\quart}{\tfrac{1}{4}}
\newcommand{\iquart}{\tfrac{\mathrm{i}}{4}}
\newcommand{\tab}{\quad\,}
\newcommand{\w}{\omega}					
\newcommand{\hphi}{\hat{\phi}}				
\newcommand{\hF}{\hat{F}}				
\newcommand{\p}{\partial}				
\newcommand{\C}{\hat{C}}				
\newcommand{\D}{\mathcal{D}}				
\newcommand{\Db}{\bar{\mathcal{D}}}			
\newcommand{\e}[2]{e_{#1}{}^{#2}}			
\newcommand{\E}[2]{E_{#1}{}^{#2}}			
\newcommand{\F}[2]{\mathcal{F}_{#1}{}^{#2}}		
\newcommand{\A}[2]{\mathcal{A}_{#1}{}^{#2}}		
\newcommand{\com}[2]{[\,#1\, ,\,#2\,]}			
\newcommand{\Chi}{\raisebox{2pt}{$\chi$}}
\newcommand{\hchi}{\raisebox{2pt}{$\hat{\chi}$}}
\newcommand{\lra}{\Longrightarrow}
\renewcommand{\O}{\Omega}				
\renewcommand{\i}{\mathrm{i}}

\begin{document}


{\tiny Institut f\"ur Theoretische Physik, Universit\"at Hannover\hfill
       Institut f\"ur Theoretische Physik, Universit\"at Hannover}
\vspace{5mm}

\setlength{\unitlength}{.8mm}

\begin{center}
 \begin{picture}(21,16)
  \multiput(10,0)(7,0){2}{\framebox(4,16){}}
  \multiput(4,0)(1,0){2}{\line(0,1){16}}
  \multiput(0,16)(5,0){2}{\line(1,0){4}}
  \put(14,8){\line(1,0){3}}
  \put(4,16){\oval(8,32)[bl]}
  \put(5,16){\oval(8,32)[br]}
 \end{picture}
\end{center}
        
\vspace{-13.5mm}

ITP--UH--21/97 \\ hep-th/9707238 \hfill July 1997


\vspace{10mm}
\begin{center}
 {\LARGE\bfseries Supergravity with a \\ Noninvertible Vierbein} \\[5mm]
 Norbert Dragon, Holger G\"unther and Ulrich Theis \\[2mm]
 \textit{Institut f\"ur Theoretische Physik, Universit\"at Hannover \\
	 Appelstra\ss{}e 2, 30167 Hannover, Germany}
\end{center}
\vspace{8mm}

\begin{center} \textbf{Abstract} \end{center}

We show that there is no off-shell Palatini formulation of minimal
supergravity. Nonetheless, we have been able to generalize the multiplet
and Lagrangians of this theory to the case of vanishing determinant of
the vierbein. Unfortunately, the requirement of regularity does not single
out a unique action.


\section{Introduction}

In \cite{dragon-reg-lagrange} it was shown that altering the field content
of general relativity to allow for noninvertible metrics gives the
Einstein-Hilbert action as the unique action describing pure gravity in four
spacetime dimensions. It was also shown, however, that this uniqueness is
lost when coupling matter to gravity and is only regained in spacetimes
with dimensions greater than or equal to nine.

It was hoped that an additional symmetry, namely supersymmetry, could
help to establish this uniqueness in lower spacetime dimensions.

As will be shown this is unfortunately not the case. We have succeeded in
formula\-ting both the multiplet and the Lagrangian of (old) minimal 
supergravity regularly in the vierbein, i.e.\ in a way that they remain well 
defined, even if $\det \e{a}{m}$ vanishes. As this is possible only by 
introducing a density of negative weight, the resulting Lagrangian is not
unique.

It must also be noted that the resulting action is not the supersymmetric
extension of the action presented in \cite{dragon-reg-lagrange}. This is a
consequence of the fact that, as will be shown in Section \ref{no palatini}, 
there is no Palatini formulation of minimal supergravity, a means which is 
substantial to \cite{dragon-reg-lagrange}. In particular we have found no 
way of describing pure supergravity.

Our approach to supergravity is along the lines of \cite{dragon-sugra},
except for the fact that we use the full BRS-transformations of the fields.

In this paper we only outline the strategy leading to our results. The
calculational details can be found in \cite{holger,ulrich}.

\section{The Algebra and the BRS-Transformations}

In this section we closely follow \cite{brandt}. We start investigating
the algebra of covariant transformations $\Delta_N$ (i.e.\ transformations
that map tensor fields on tensor fields):
 \begin{equation} \label{algebra}
  \com{\Delta_M}{\Delta_N}\, T := \big( \Delta_M \Delta_N - (-)^{|M||N|}
	\Delta_N \Delta_M \big) T = \F{MN}{P} \Delta_P T\, ,
 \end{equation}
where the capital latin indices denote collectively the following operators:
 \begin{center}
  \begin{tabular}{rl}
   $\D_a$ : & covariant spacetime derivatives, \\
   $\D_\al$, $\Db_{\ad}$ : & supersymmetry transformations \\
   & (spinor derivatives), \\
   $\ell_{[ab]}$ : & Lorentz spin transformations.
  \end{tabular}
 \end{center}
The algebra splits into the (super) spacetime transformations
($A \in \{ a, \al, \ad \}$) and the Lorentz spin transformations ($[ab]$):
 \begin{equation} \label{algebra2} \begin{split}
  \com{\D_A}{\D_B} & = T_{AB}{}^C \D_C + \half R_{AB}{}^{[ab]}
	\ell_{[ab]}\, , \\[4pt]
  \com{\ell_{[ab]}}{\D_A} & = - G_{[ab]A}{}^B\, \D_B\, , \\[4pt]
  \com{\ell_{ab}}{\ell_{cd}} & = \eta_{ad}\, \ell_{bc} - \eta_{ac}\,
	\ell_{bd} - \eta_{bd}\, \ell_{ac} + \eta_{bc}\, \ell_{ad}\, .
 \end{split} \end{equation}

The partial derivative acting on tensors defines the connections,
 \begin{equation}  \label{partial}
  \p_m T = \A{m}{N}\! \Delta_N T\, ,
 \end{equation}
which are denoted by
 \begin{center}
  \begin{tabular}{rl}
   $\e{m}{a}$ : & inverse vierbein\footnotemark, \\
   $\half \psi_m{}^\al$, $\half \bpsi_m{}^{\ad}$ : & Rarita-Schwinger
	field, \\
   $-\w_m{}^{[ab]}$ : & spin connection.
  \end{tabular}
 \end{center}
\footnotetext{Our choice of what is called vierbein and what inverse
	      vierbein is non-standard. It is motivated by the fact
	      that $\e{a}{m}$ is the field appearing in typical kinetic
	      terms, e.g.\ $\eta^{ab} \e{a}{m} \e{b}{n} \p_m \phi\, \p_n
	      \phi^*$. For the same reason the inverse metric was considered
	      fundamental in \cite{dragon-reg-lagrange}.}

The BRS-transformations are defined similarly, replacing the connections
by ghosts
 \begin{equation} \label{brs1}
    s T = C^N\! \Delta_N T\, .
 \end{equation} 
By requiring that $\com{s}{\p_n} = s^2 = \com{\p_n}{\p_m} = 0$ hold 
on tensor fields, one can deduce the BRS-transformations of the connections 
and ghosts
 \begin{align}
   s \A{m}{P} & = \p_m C^P + C^N \A{m}{M} \F{MN}{P}\, , \\[4pt]
	s C^P & = \half (-)^{|N|} C^N C^M \F{MN}{P}\, ,
 \end{align}
and express some of the structure functions in terms of connections
($\mu \in \{ \al, \ad, [ab] \}$):
 \begin{equation} \label{identifikation} \begin{split}
  \F{ab}{N} = - \e{a}{m} \e{b}{n} \big( & \p_m \A{n}{N} - \p_n \A{m}{N} +
	\A{n}{\mu} \A{m}{\nu} \F{\mu\nu}{N} \\*
  & - \A{n}{\mu} \e{m}{c} \F{\mu c}{N} + \A{m}{\mu} \e{n}{c} \F{\mu c}{N}
	\big)\, .
 \end{split} \end{equation}

Notice that here and in almost all other intermediate steps we do not
question the invertibility of the vierbein. Our final results, however,
will remain well defined even if it is noninvertible. These could of course
be stated without recourse to the intermediate steps, which are therefore
to be regarded as merely motivating.

\section{Impossibility of a Palatini formulation} \label{no palatini}

Minimal supergravity is characterized by a set of constraints on the
algebra \eqref{algebra}, and one of these is usually taken to
be $T_{ab}{}^c=0$. Omitting this one (see discussion below), equation
\eqref{identifikation} reads in particular:
 \begin{equation} \label{torsion}
  T_{ab}{}^c = \e{a}{m} \e{b}{n} (\p_n \e{m}{c} - \p_m \e{n}{c}) + \ihalf
	(\psi_a \si^c \bpsi_b - \psi_b \si^c \bpsi_a) + \w_{ba}{}^c -
	\w_{ab}{}^c\, .
 \end{equation}
Using the antisymmetry of $\w_a{}^{bc}$ in its last two indices, one finds
 \begin{equation} \label{spinkonfestgelegt} \begin{split}
  \w_{abc} & = \half \big( \e{a}{m} \e{b}{n} \eta_{cd} + \e{c}{m}
	\e{a}{n} \eta_{bd} - \e{b}{m} \e{c}{n} \eta_{ad} \big) \big( \p_n
	\e{m}{d} - \p_m \e{n}{d} \big) \\*
  & \tab + \iquart \big( \psi_a \si_c \bpsi_b - \psi_b
	\si_c \bpsi_a + \psi_c \si_b \bpsi_a - \psi_a \si_b \bpsi_c
	- \psi_b \si_a \bpsi_c + \psi_c \si_a \bpsi_b \big) \\*
  & \tab - \half (T_{abc} - T_{acb} - T_{bca}) \\
  & =: \O_{abc}(e, \psi) - \half (T_{abc} - T_{acb} - T_{bca})\, .
 \end{split} \end{equation}

Here one sees that constraining the torsion $T_{ab}{}^c$ to vanish would fix
the spin connection in terms of the vierbein, the inverse vierbein and the
Rarita-Schwinger field. This function, denoted by $\O_a{}^{bc}$, is
certainly not regular at vanishing determinant of $\e{a}{m}$, hence it must
not appear in the transformation laws and the Lagrangian, as these are
required to remain well defined if $\det \e{a}{m} = 0$. The analogous
problem in general relativity was solved in \cite{dragon-reg-lagrange} by
taking recourse to a Palatini formulation, in which the connection is treated
as an independent field whose equations of motion give the above
identification.

In this section we show why this is not possible in minimal supergravity,
while in the following sections we proceed to show how regular
transformations and Lagrangians can be obtained, nonetheless. 

It seems that abandoning the constraint $T_{ab}{}^c = 0$ should be sufficient
for having a Palatini formulation. To see that it is not, consider the change
in the algebra \eqref{algebra2} induced by
 \begin{equation} \label{new-basis}
  \D_a' := \D_a + \quart (T_a{}^{bc} - T_a{}^{cb} - T^{bc}{}_a)\,
	\ell_{bc}\, ,\quad \D_\al' = \D_\al\, 
        ,\quad \ell_{ab}' = \ell_{ab}\, .
 \end{equation}
It leads to
 \begin{equation}
  \com{\D_A'}{\D_B'} = T'_{AB}{}^C \D_C' + \half R'_{AB}{}^{ab}
	\ell_{ab}\, ,
 \end{equation}
where the primed torsions are given by (the primed curvatures are of no
concern to us in the following)
 \begin{equation} \label{neue Torsion}
  T'_{ab}{}^c = 0\, ,\quad T'_{a\be}{}^\ga = T_{a\be}{}^\ga - \quart
	(T_{abc} - T_{acb} - T_{bca}) \si^{bc}{}_\be{}^\ga\, .
 \end{equation}
No other torsions are changed, in particular
 \begin{equation} \label{no-change}
   T'_{ab}{}^{\ga} = T_{ab}{}^{\ga}\, .
 \end{equation}

Expressing \eqref{partial} and \eqref{brs1} in the new basis, one also 
obtains a change in the ghosts and connections:
 \begin{gather}
  s T = C^N\! \Delta_N T = C^{'N}\! \Delta'_N T \notag \\
  \lra C^{'A} = C^A\, ,\quad C^{'[ab]} = C^{[ab]} + \half C^a (T_a{}^{bc}
	- T_a{}^{cb} - T^{bc}{}_a) \label{neue BRS} \\[4pt]
  \p_m T = \A{m}{N} \Delta_N T = \mathcal{A}'_m{}^N \Delta'_N
	T \notag \\
  \lra \mathcal{A}'_m{}^A = \A{m}{A}\, ,\quad \w'_m{}^{[bc]} =
	\w_m{}^{[bc]} + \half \e{m}{a} (T_a{}^{bc} - T_a{}^{cb} -
	T^{bc}{}_a)\, .
 \end{gather}

One sees that this change of basis is equivalent to imposing the constraint
$T_{ab}{}^c=0$. All this is not unlike the case of general relativity. What
has been said up to now is hardly more than an elaborate way of stating that
two connections differ by a tensor.

The distinction comes when one starts to write down Lagrangians. In general
relativity one can use the Ricci scalar expressed in terms of the
independent or dependent spin connection according to the algebra used.
Supergravity, however, is more restrictive. The Ricci scalar enters the
action through the $\D^2 M$ term in the Lagrangian (compare with
\eqref{normal-lag}: $\D^2$ comes from the $\hF$ term and $M$, the auxiliary
scalar in the supergravity multiplet, from the chiral projector). From the
Bianchi identities (we use the conventions of \cite{holger, ulrich}) one
knows:
 \begin{equation} \label{new-ricci}
  \D^\al \D_\al M = \tfrac{8}{3} \si^{ab}{}_\ga{}^\al \D_\al T_{ab}{}^\ga\, .
 \end{equation}
But the derivatives as well as the torsions which appear in this equation are
left invariant by the above change of basis \eqref{new-basis},
\eqref{no-change}, so that one also has
 \begin{equation} \label{old-ricci}
  \D^\al \D_\al M = \tfrac{8}{3} \si^{ab}{}_\ga{}^\al \D'_\al T'_{ab}
	{}^\ga\, .
 \end{equation}
But this means that this part of the Lagrangian is not altered at all by the
change of basis and therefore a Palatini formulation does not exist in minimal
supergravity. One can understand this result in the following way:

A crucial difference between general relativity and supergravity is the fact
that in the latter theory all curvatures are determined by the torsions
\cite{dragon-bianchi}. But this means that changing the constraints on the
torsions will also effect the curvatures and in particular the Ricci scalar.
Indeed, one finds that abandoning $T_{ab}{}^c=0$ allows to express the Ricci
scalar in terms of an independent spin connection, but on the other hand one
is forced to add additional torsion terms to the action. If now
\eqref{torsion} is inserted for $T_{ab}{}^c$, all terms containing the 
independent spin connection ($\w$) cancel and one is left with the Ricci 
scalar expressed in the dependent spin connection ($\O$).

In formulae (denoting the torsion terms by $f(T)$):
 \begin{equation}
  R(\O)\; \stackrel{T\neq 0}{\lra}\; R(\w) + f(T) = R(\w) + f(T(\w,\, \O))
	= R(\O)\, .
 \end{equation}
But this is just the above statement: Abandoning the constraint
$T_{ab}{}^c=0$ does not alter the Ricci scalar part of the supergravity 
Lagrangian. Explicit calculation shows that the entire Lagrangian is
unaffected by this change of basis of the algebra.

One also sees that it is possible to express the transformation laws of the
fields entirely in terms of the dependent spin connection as the change of
basis leaves the BRS-operator invariant.

\section{Regular Transformations}

Looking at the explicit form of the BRS-transformations (see e.g.\ 
\cite{dragon-sugra}) one sees that these are not regular expressions in
$\e{a}{m}$, due to covariant derivatives appearing e.g.\ in the
transformation law of the Rarita-Schwinger field. These contain the spin
connection, which, as was demonstrated, is not regular. However, one also
sees that terms containing the spin connection are the only singular
terms, if one uses $\{ \e{a}{m},\, \A{a}{\mu} = \e{a}{m} \A{m}{\mu} \}$
instead of the $\A{m}{N}$ as fundamental fields.

To obtain a regular multiplet we introduce a vierbein density as a
fundamental field:
 \begin{equation} \label{vbdensity} \begin{split}
  \E{a}{m} & := e^{-q} \e{a}{m}\, ,\quad \E{m}{a} = e^{q} \e{m}{a}\, , \\
  \e{a}{m} & = E^{\kappa} \E{a}{m}\, ,\quad \e{m}{a} = E^{-\kappa}
	\E{m}{a}\, ,
 \end{split} \end{equation}
where $e=\det \e{a}{m}$, $E=\det \E{a}{m}$, $\kappa= q/(1-4q)$ and $q$ is a
real number to be specified later. The above expressions can obviously not 
be used to define the new fields at vanishing $e$. They allow, however, to 
calculate the BRS-transforma\-tions of these fields which remain well defined 
even at $e=0$ and serve as the proper definition of $\E{a}{m}$.
\eqref{vbdensity} then shows how to regain the old vierbein at nonvanishing 
determinant.

We can now examine the singular terms contained in the spin connection.
These are all of the form
 \begin{equation}
  \e{b}{m} \e{a}{n} \p_n \e{m}{c} = - \e{m}{c} \e{a}{n} \p_n \e{b}{m}
 \end{equation}
and read, expressed in the new fields:
 \begin{equation} \label{new}
  \E{m}{c} \E{a}{n} \p_n (E^{\kappa} \E{b}{m}) = E^{\kappa} \E{m}{c}
	\E{a}{n} \p_n \E{b}{m} + \kappa \de_b^c \E{a}{n} E^{\kappa} \E{m}{d}
	\p_n \E{d}{m}\, .
 \end{equation}
The inverse of a matrix is given by a polynomial expression in its matrix
elements (called the minor) devided by its determinant. This means that
$E \E{m}{a}$ is a regular expression in $\E{a}{m}$ even though $\E{m}{a}$ is
not. In particular \eqref{new} is regular in the $\E{a}{m}$ if only $\kappa$
is chosen to be greater than or equal to one. It is polynomial if $\kappa$
is a natural number.

Thus it is immediately obvious that by densitizing the vierbein all
singularities can be removed from the transformation laws of the ghosts and
the connections and only the transformation of the vierbein density itself
remains to be checked. One finds
 \begin{equation} \begin{split}
  s \E{a}{m} & = \frac{\kappa}{1 + 4 \kappa} \big( E^{\kappa} \E{b}{n}
	\p_n C^b + C^N\! \A{b}{M} \F{MN}{b} \big) \E{a}{m} \\
  & \tab - E^{\kappa} \E{b}{m} \E{a}{n} \p_n C^b - \E{b}{m} C^N\!
	\A{a}{M} \F{MN}{b}\, ,
 \end{split} \end{equation}
which again is regular. 

The restriction $\kappa \geq 1$ translates to the requirement that $q$ be
from the half-open interval $[\tfrac{1}{5}\, ,\, \quart)$. $\kappa$ is a
natural number if $q$ has the form $q=n/(1+4n)$ with $n$ natural.

If one now calculates the transformation of $E$, one finds
 \begin{align}
  s E & = (4q-1) \big[ \p_n (E^{\kappa} \E{a}{n} C^a) + \i \big( (C - \half
	C^a \psi_a) \si^b \bpsi_b + \psi_b \si^b (\bar{C} - \half C^a
	\bpsi_a) \big) \big] E \notag \\*
      & \tab + E^{\kappa} \E{a}{n} C^a \p_n E \notag \\
      & = (4q-1) \big[ \p_n \C^n + \i (\C \si^b \bpsi_b + \psi_b \si^b
	\Hat{\Bar{C}}) \big] E + \C^n \p_n E\, , \label{dichte-term}
 \end{align}
where
 \begin{equation}
  \C^n := E^{\kappa} \E{a}{n} C^a\, ,\quad \C^\al := C^\al - \half C^a
	\psi_a{}^\al\, .
 \end{equation}
Recognizing that the first term is the supersymmetric generalization of
the cha\-racteristic term in the transformations of densities, one sees that
$E$ transforms as a density of weight $4q-1$, which is negative in the
allowed range of $q$.

The results of this section show that one can obtain a regular second order 
formulation of general relativity if one uses vierbein densities as
fundamental fields. One cannot achieve this in a metric/affine
connection formulation by introducing (inverse) metric densities. 

\section{The Lagrangian}

In minimal supergravity one constructs Lagrangians from chiral fields. This
cannot at once be generalized to our case. The problem is that a Lagrangian
has to transform as a density of weight one, which is usually achieved by
choosing a scalar times the determinant of the inverse vierbein
($\det \e{m}{a}$) as the Lagrangian. This, of course, is not possible here,
as we want to investigate theories in which $\e{m}{a}$ need not exist.

The remedy is to investigate not ordinary chiral fields, but chiral 
densities, i.e.\ to construct a Lagrangian from fields that themselves
contribute weight, so that no explicit determinants are needed.
As then the weight of the Lagrangian comes from matter fields we will
however not be able to describe pure supergravity.

$(\hphi\, ,\hchi_\al\, ,\hF)$ being the components of a chiral multiplet,
we define
 \begin{equation} \label{tensordichten}
  \phi := e^{-p} \hphi\, ,\quad \Chi_\al := e^{-p} \hchi_\al\, ,\quad F :=
	e^{-p} \hF\, ,\qquad e = E^{1/(1-4q)}\, (=\det \e{a}{m})\, .
 \end{equation}
``Defining'' has, as in \eqref{vbdensity}, to be regarded as tongue-in-cheek.
\eqref{tensordichten} motivates the (regular) transformations of these fields
which in turn are used to define them properly. These transformations will
however not be explicitely stated here as they would introduce
technicalities not necessary to understand the following. It shall suffice
to mention the instances in which chiral densities require treatment
different from chiral fields.

Chiral fields are characterized by the equation $\Db_{\ad} \hphi = 0$,
which is equivalent to the requirement that their transformation must not
contain a ghost $\bar{C}^{\ad}$. Looking at the characteristic term for the
transformations of densities \eqref{vbdensity}, one sees that this 
condition has to be altered. It turns out to be sufficient to subtract 
the extra term. Introducing the new operators
 \begin{equation}
  \de_\al := \Big[\, \frac{\p}{\p \C^\al}\, ,\, s\, \Big] - \i (\si^a
	\bpsi_a)_\al\, \mathcal{W} \, ,\quad
  \bar{\de}_{\ad} := - \Big[\, \frac{\p}{\p \Hat{\Bar{C}}^{\ad}}\, ,\, s\,
	\Big] + \i (\psi_a \si^a)_{\ad}\, \mathcal{W} \, ,
 \end{equation}
where $\mathcal{W}$ assigns to a field its weight (e.g.\ $\mathcal{W}(\phi) =
p \phi$), one can write the defining 
equation for chiral densities as $\bar{\de}_{\ad} \phi = 0$. Also the
equations which give the higher components of chiral fields can be
generalized to chiral densities:
 \begin{equation}
  \Chi_\al = \de_\al \phi\, ,\quad F = - \quart \de^2 \phi\, .
 \end{equation}
The corresponding equations hold for antichiral densities\footnote{The
operators $\de_\al$ and $\bar{\de}_{\ad}$ should not be regarded as
generalising the spinor derivatives to act on tensor densities. While they
reduce to $\D_\al$ and $\Db_{\ad}$ when acting on tensors, they do not
satisfy the algebra \eqref{algebra} on tensor densities.}.

As $\de_\al$ and $\bar{\de}_{\ad}$ satisfy the (graded) Leibniz rule a product
of chiral densities is again a chiral density, the weight being the sum of
the weights of its factors.

The chiral projector known from minimal supergravity generalises to chiral
densities, i.e.\ the operator $(\bar{\de}^2 - 2M)$ acting on a product of
chiral and antichiral densities gives a chiral density. The weight of this
density is again the sum of the weights of the factors. The proof is
straightforward, but will be omitted here as we have not discussed the
transformation laws of the auxiliary fields.

With these results it is now easy to construct Lagrangians. From minimal
supergravity it is known that
 \begin{equation} \label{normal-lag}
  e^{-1} \big( \hF + \ihalf \bpsi_a \bsi^a \hchi + \tfrac{3}{2} M^\ast
	\hphi - \bpsi_a \bsi^{ab} \bpsi_b\, \hphi \big)\, ,\quad e =
	E^{1/(1-4q)}\, (=\det \e{a}{m})
 \end{equation}
transforms into a total derivative. But by definition
 \begin{equation} \begin{split}
  & F + \ihalf \bpsi_a \bsi^a \Chi + \tfrac{3}{2} M^\ast \phi - \bpsi_a
	\bsi^{ab} \bpsi_b\, \phi \\
  = \big(& - \quart \de^2 + \ihalf \bpsi_a \bsi^a \de + \tfrac{3}{2} M^\ast
	- \bpsi_a \bsi^{ab} \bpsi_b \big) \phi
 \end{split} \end{equation}
has the same transformation law and therefore gives a Lagrangian, if the 
chiral density $\phi$ has weight one, without an inverse vierbein appearing
explicitly. Since we know how to construct chiral densities from chiral 
and antichiral densities we can immediately give the formula for the most 
general Lagrangian we are able to build:
 \begin{equation} \begin{split}
  \mathcal{L} & = \big( - \quart \de^2 + \ihalf \bpsi_a \bsi^a \de +
	\tfrac{3}{2} M^\ast - \bpsi_a \bsi^{ab} \bpsi_b \big) \big[
	(\bar{\de}^2 - 2M)\, K(\phi, \phi^\ast) + g(\phi) \big] \\
	      & \tab + \text{h.c.}\, .
 \end{split} \end{equation}
Here $K$ is a polynonial of chiral and antichiral densities, while $g$ is a
polynonial of chiral densities only. In both cases each individual monomial
must be a density of weight one.

One has to remember, however, that the theory contains a chiral field of 
negative weight, namely $E$, \eqref{vbdensity}. This means that one can take 
an arbitrary polynomial in elementary chiral densities and multiply its
terms with appropriate powers of $E$ to obtain again a chiral density of
weight one. This in turn demonstrates that one does not obtain a unique 
action from symmetry and regularity alone. Indeed, one sees that to every
action of minimal supergravity coupled to chiral matter there exists a
corresponding regular action.


\begin{thebibliography}{99}

 \bibitem{dragon-reg-lagrange}
	N.~Dragon,
	\emph{Regular Gravitational Lagrangians},
	Phys.~Lett.~B \textbf{276} (1992) 31.

 \bibitem{dragon-sugra}
	N.~Dragon, U.~Ellwanger, M.~G.~Schmidt,
	\emph{Supersymmetry and Supergravity},
	Prog.~Particle Nucl.~Phys.~\textbf{18} (1987) 1.

 \bibitem{holger}
	H.~G\"unther,
	\emph{Supergravitation mit nichtinvertierbaren Vierbeinen},
	diploma thesis, University of Hannover (1996), unpublished.

 \bibitem{ulrich}
	U.~Theis,
	\emph{Palatini-Formulierung der Supergravitation},
	diploma thesis, University of Hannover (1996), unpublished.

 \bibitem{brandt}
	F.~Brandt,
	\emph{Structure of BRS Invariant Local Functionals},
	NIKHEF Preprint NIKHEF-H-93-21, hep-th/9310123 (1993).

 \bibitem{dragon-bianchi}
	N.~Dragon,
	\emph{Torsion and Curvature in Extended Supergravity},
	Z.~Physik C, Particles and Fields \textbf{2} (1979) 29.

\end{thebibliography}
\end{document}